\documentclass[prl,twocolumn,Letters, superscriptaddress]{revtex4}

\usepackage{graphicx}
\newcommand{\beq}{\begin{equation}}
\newcommand{\eeq}{\end{equation}}
\newcommand{\beqa}{\begin{eqnarray}}
\newcommand{\eeqa}{\end{eqnarray}}

\newcommand{\ket}[1]{\mbox{$ | #1 \rangle $}}
\newcommand{\bra}[1]{\mbox{$ \langle #1 | $}}

\def\half{\frac{1}{2}}
\def\opone{\leavevmode\hbox{\small1\normalsize\kern-.33em1}}

\begin{document}

\title{How far can one send a photon ?}

\author{Nicolas Gisin \\
\it \small   Group of Applied Physics, University of Geneva, 1211 Geneva 4,    Switzerland}

\date{\small \today}

\begin{abstract}
The answer to the question {\it How far can one send a photon?} depends heavily on what one means by {\it a photon} and on what one intends to do with that photon. For direct quantum communication the limit is of about 500 km. For terrestrial quantum communication, near future technologies based on quantum teleportation and quantum memories will soon enable quantum repeaters that will turn the development of a world-wide-quantum-web (WWQW) into a (highly non-trivial) engineering problem. For Device Independent Quantum Information Processing, near future qubit amplifiers (i.e. probabilistic heralded amplification of the probability amplitude of presence of photonic qubits) will soon allow demonstrations over a few tens of km.
\end{abstract}

\maketitle

\section{Introduction}
{\it How far can one send a photon?} Well, some day one will be able to store the photon in a quantum memory and send it by airplane to the other side of Earth. But this is a distant future vision, hence, let’s rephrase our question. {\it How far can one send a photon with today’s, or near future, technologies?} Well, I am afraid that it still depends. There will soon be quantum communications between low orbit satellites and Earth, i.e. over several hundreds of kilometers. And if one looks at light from distant stars or galaxies, then one analyzes photons having travelled astronomical distances. But that’s a bit cheating, we didn’t send those photons. So, let’s rephrase again our question. {\it How far can one send a photon on Earth using today’s, or near future, technologies?} Hum, it depends.

In classical optical communications one routinely sends light pulses across oceans, e.g. from France to the US over a single and continuous optical fiber about 5000 kilometers long. So, do photons propagate all the way over the Atlantic? Assuming a mean loss of 0.2 dB/km the total loss is 1000 dB, i.e. a transmission of $10^{-100}$. So, even if the initial light pulse has millions of photons (let’s say, 1 ns of a mW, i.e. $10^{-12}$ Joule, hence about $10^7$ photons), there is a negligible chance that a photon from the initial pulse makes it from France to the US. But classical optical communication works: about every 50 km, when only 10\% of the photons remain, a laser mechanism amplifies the light pulse back to its initial state, up to some unavoidable-but-tolerable noise. This is very nice and effective. But can one say that some photons make the entire journey? Probably not, though it depends what a photon is. If one thinks of a photon produced by stimulated emission as a new photon, then the chance that any photon leaving France arrives in the US is vanishingly small. But photons are excitations of an optical mode (e.g. the single mode of optical fibers), hence the question is not well defined. Hence, let’s be more precise. {\it How far can one send a bit of information encoded in a single photon, on Earth, using today’s, or near future, technologies?} This excludes classical optical communication, as a bit encoded in a single-photon gets lost when the photon passes a classical amplifier.

\section{Direct Quantum Communication}
On Earth, the best choice of communication channel is an optical fiber, no doubt about this. Today’s best fibers have losses close to 0.16 dB/km, hence an optimistic loss for near future fibers is around 0.15 dB/km – the probability that a photon makes it through 500 km is thus $10^{-7.5}$. This is low, but not unreasonably low. Indeed, if the rate at the transmitter is around 10 GHz, then, on average, about hundred photons arrive at the receiver every second. And today there are single-photon detectors at telecom wavelengths (i.e. compatible with the highest transmission of telecom optical fibers) close to 100\% efficiency and with remarkable low dark counts \cite{snspdMarsili}. The longest distance single-photons, more precisely pseudo-single-photons, i.e. weak pulses with less than one photon per pulse, have been sent and detected is, I believe, the recent QKD demonstration by my colleague Hugo Zbinden's group over 307 km of ultra-low-loss fibers from Corning, operating at a source clock rate close to 1 GHz \cite{QKD307km}. But beware, be difference between 300 and 500 km is huge because the transmission probability per photon goes down exponentially. Consequently 500 km is still out of reach. At best one could go for 100 GHz sources, i.e. a gain of 20 dB, improve the detectors from 20\% (as in \cite{QKD307km}) to 100\% (+7 dB), lower the fiber loss from 0.16dB/km down to 0.15dB/km (a gain of +3dB over the 300 km), use true single-photons instead of the 0.5 photon per pulse (+3dB) and ... that’s it; there exist no further straightforward improvements! Hence a total potential gain of 33 dB corresponding to an additional length of 33/0.15=220 km for a total distance just above 500 km \cite{QCommReviewRob}. I would be truly surprised if anyone demonstrates a longer distance during my lifetime (which I consider as the near future).

So, can one do better? Well, it depends. First, it depends on the minimal rate, let's say a few tens of bits per second. But more interestingly, there are, actually, at least two intriguing possibilities to beat the direct quantum communication upper bound of 500 km: heralded probabilistic amplification of photonic qubits, and quantum repeaters.

\section{Quantum repeaters}
Let us first analyze the second possibility, i.e. quantum repeaters \cite{BriegelRepeater}. The basic ingredient is quantum teleportation, this marvelous process that exploits entanglement as a quantum communication channel \cite{Qtelep, TelepReview}. Once two qubits are entangled, this can be used to teleport the quantum state of a third qubit from one side to the other, independently of the distance separating the two entangled qubits. Assume all three qubits are photonic qubits, e.g. polarization qubits or time-bin qubits \cite{Brendel1999,Tittel-Weihs}. Since the receiving qubit and photon are indistinguishable from the initial qubit and photon, one may argue that this corresponds to sending one photon from the location of the initial qubit all the way to the receiver. But through which trajectory? Well, in quantum teleportation there is no trajectory, nothing in our 3-dimension space (it all happens in Hilbert space). So, if one likes to include this process in our list of answers to our basic question {\it How far can one send a photon?}, then one could argue that it is the straight line that determines the relevant distance. Accordingly, if one teleports a photonic qubit from Europe to Australia, the distance is not half the circumference of Earth, but its diameter. But this wouldn't do justice to the scientific and technological challenge. Indeed, before teleporting anything one first has to entangle the two end points and this is done by first defining a trajectory relating the end points and next dividing it into many relatively short sections. Note that this is a bit similar to the case of classical optical communication using a laser mechanism to re-amplify the optical signal about every 50 km, as discussed above; also the question whether the photon at the end point is the same as the initial one is a bit similar. So, the very way quantum teleportation on a long terrestrial distance will someday be realized defines a natural notion of distance.

Ok, now we have our long distance cut into many sections, see Fig. 1. Note that if one would work in series, that is first teleport a qubit from the first point to the second, next from the second to the third, and so on until the final receiving point, then the process would be exponentially inefficient in the total distance. Indeed, one would first need to entangle the first point to the second and perform a first teleportation, next to entangle the second point with a third one and perform a second teleportation, and so on. But to entangle two points requires sending photons between these two points (either from a middle station to both points, or from one point to the other), and this is as inefficient as sending directly the photon carrying the qubit \cite{QCommReviewRob}. 

\begin{figure}
\centering
\includegraphics[width=8cm]{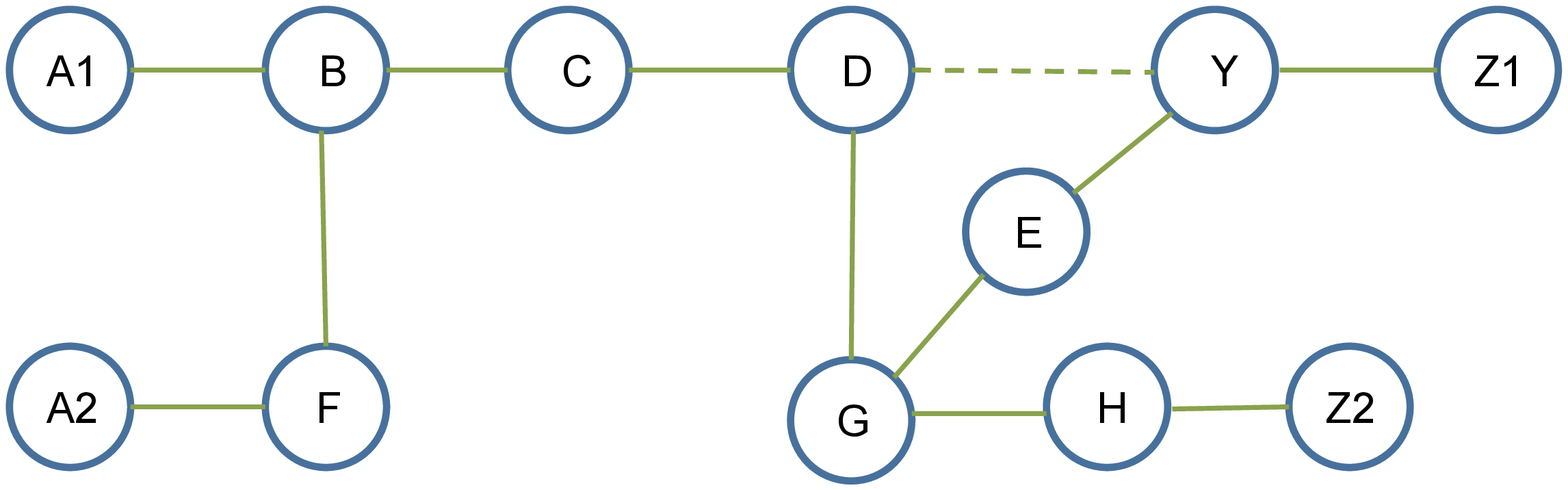}
\caption{Example of a small quantum network with nodes $A_j,B,..,Y,Z_j$, $j=1,2$. Nodes $B,...,Y$ must hold  as many quantum memories as there are links to that node. Each link exploits quantum teleportation.}\label{fig1}
\end{figure}

Hence, operating in series in not more efficient than direct communication (and actually much less efficient because of the many teleportation processes). One needs to operate in parallel: trying to entangle every pair of nearby nodes at the same time. But again, the probability that all these entanglement distributions work all at the same time is exponentially inefficient. Hence, the only way to exploit teleportation to improve the overall efficiency is to allow that some pairs of nearby points get entangled before others and wait for the others. 

But how can one say to a photon {\it Please, wait a while}? Well, that requires synchronization and that can be done using quantum memories. In summary, without quantum memories, there is no quantum repeater. 

Moreover, the quantum memories must have some further quality, ideally one should be able to send in a photon (write) and release it (read) on demand and know when the memory is loaded, i.e. when the qubit is stored. Actually, the exact requirements are subtle and depend on the exact protocol, we will not enter this discussion here (and it is likely that the best protocol has not yet been invented). Let me just emphasize that today, somewhat surprisingly, the best known protocol doesn't use photon pairs, but uses single-photons \cite{Sangouard2007}.

There are at least 5 specifications for quantum memories that are crucial and independent of any protocol:
\begin{enumerate}
\item Efficiency, i.e. the probability that an incoming photon is stored and properly released at the appropriate time. Note that inefficiency is equivalent to additional loss, hence implies shorter sections.
\item Fidelity, i.e. the overlap between the states of the in-qubit and the out-qubit, should be close to one. This is a much discussed parameter, but in practice, those photons that are not lost have a very high fidelity. Hence, thanks to the natural post-selection of those photons that make it to the detector, quantum communication with discrete variables has some built-in error filtering.
\item Memory time which determines the size of the quantum network that can be synchronized. At first, only synchronization between nearby points is required, however once next neighbors are entangled, synchronization over longer and longer distances is required. Hence, the quantum memory time defines the longest distance and our question becomes {\it How long can one hold a photonic qubit in a quantum memory?} In practice one should add a factor of ten as margin to allow for time to repeat the entanglement distribution process many times until success. Today, the longest time for read-write quantum memories is only about one millisecond\footnote{There are quantum memories with longer storage times, however they don't allow to store an incoming photon; either they generate themselves photons which are entangled with the quantum memory, hence there are not read-write quantum memories, but read-only memories \cite{Kuzmich100ms}, or they don't have any input-output \cite{Sellars6h}.} \cite{Jobez_ms,chinois_ms} insufficient to beat direct communication. But steady progress allows one to be optimistic. For single-photon read-write quantum memories, second long storage times are on the horizon. Recall that {\it storage time} is defined as the time of the exponential decay of the memory efficiency; recall that for single-photon memories, those photons that come out of the memory do so with high fidelities, in particular with fidelities above the optimal cloning threshold.
\item There is a fourth very important specification: the bandwidth. Indeed, quantum memories useful for quantum communication should be able to store photons with a bandwidth of at least tens of MHz. A few GHz would be better. Today, the best quantum memory from this point of view has been demonstrated in Calgary-Canada \cite{WT_1GHzQM}.
\item A final specification required for quantum memories as building blocks for quantum repeaters is the ability to store multiple of photonic qubits, multiplexed either in time \cite{UsmaniMultiMode}, space or frequency \cite{WTfreq}. Indeed, Once a memory has been used, i.e. stores a qubit, one has to wait for information from other nodes in the network before either releasing the qubit or cleaning out the quantum memory so that it is again ready for a fresh qubit. And this communication time multiplied by the success probability of some entanglement distribution or some Bell-state measurement can be rather large. Accordingly, without a multimode capacity, a quantum memory would not be able to make large quantum networks realistic \cite{SimonMultiMode}.
\end{enumerate}

\section{Heralded amplifier}
Let us now analyze the other possibility to overcome the 500 km hard limit for direct long-distance quantum communication mentioned above. The fact is that one can sort of “amplify” a photon \cite{T.C.Ralph2009}! This fact is, by itself, surprising and fascinating, hence deserves to be researched by physicists. There is no contradiction with the no-cloning theorem as what gets amplified is only the probability amplitude that the photon is not lost. And the process is probabilistic, that is, it doesn’t work each time. But when it works, there is a heralding signal. 

Imagine a box with an incoming and outgoing optical fiber, see Fig. 2. There is also a synchronization electronic input that informs the box at which times it may expect a photon. At those times, there is some probability that an electronic signal heralds the presence of a photon in the outgoing fiber. More precisely, the signal heralds that there is a large chance that there is a photon in the outgoing fiber, a photon identical to the one that was expected in the incoming fiber. Clearly, such a box is useful only if the chance of there being an outgoing photon, when the heralding signal is on, is larger than the chance there was a photon at the input. Finally, if that initial photon carried a qubit (e.g. in polarization or a time-bin qubit), then the outgoing photon would carry the same qubit. 

\begin{figure}
\centering
\includegraphics[width=8cm]{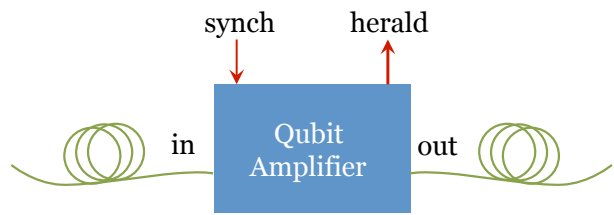}
\caption{A qubit amplifier is a box that has an input optical fiber ``in'' and an output fiber ``out'', a synchronization electronic input and an electronic heralding signal. The synchronization signal informs the box that there is a chance that a photonic qubit arrives at the input fiber. Sometimes the heralding signal informs the user that there is a larger probability that the photonic qubit is present at the output fiber.}\label{fig2}
\end{figure}

Below I briefly explain how such a box works, but first let’s ask ourselves whether such a “photon-amplifier” box helps sending photons over long distances. Well, it depends (obviously). But this time it depends on what you want to do with the photon. If the purpose is merely to detect the photon at the receiving side, then “photon-amplifiers” are of no use. However, in modern quantum communication there is another task that requires to know, with pretty high probability, that a photon is arriving before choosing in which basis to measure it. 

This task is called Device-Independent Quantum Information Processing (DIQIP), a truly remarkable subfield of Quantum Information Science. DIQIP rests on the fact that some quantum correlations can't  be mimicked by pre-established shared randomness and local processing \cite{DIQIPScarani}. Such correlations are called non-local and violate some Bell inequalities \cite{RMP-NL-Brunner14,Qchance}. They gave rise to heated debates, first between the founding father of quantum theory, next between the vast majority of physicists from the "shut up and calculate" school and the small community of physicists and philosophers interested in the foundation of quantum physics. But nowadays these non-local correlations are recognized as a resource to perform seemingly impossible tasks. Intuitively, the power of non-local correlation comes from the mere fact that if no local variables can mimic them, then neither can an adversary. Let me emphasize how remarkable this is: the mere observation that some correlation violates some Bell inequality suffices to guarantee that this correlation contains privacy \cite{AcinSecrEnt}, whether private randomness or a cryptographic key (i.e. private randomness shared between several partners). There is no need for the quantum formalism, no need for Hilbert spaces, self-adjoint operators and state-vectors: the mere observation of the correlation suffices: this is pure “quantum magic” (quantum because, according to today’s physics only quantum theory predicts the existence of non-local correlations). But for this to work one needs to take into account all trials, also those in which the photon never made it to the receiver. Indeed, if too many photons are missed, then the correlation between the remaining photons can be mimicked by local variables: local variables (plus local processing) can take advantage of the possibility to mimic lost photons. The precise threshold on the ratio of photons that have to be detected for DIQIP to be possible is still not completely known, but it is known that it has to be pretty high, somewhere between 70 and 90\% \cite{DIQIPScarani,RMP-NL-Brunner14}.

I bet that {\it photon amplifiers} will used in the first demonstrations of DIQIP over tens of km. So, how far can one go with photon-amplifiers? Well, with perfect photon-amplifiers one can go as far as for single-photons, i.e. about 500 km without quantum repeaters. Indeed, if one can detect one photon per second, then one can also amplify all the photon probability amplitudes and an ideal photon-amplifier box will herald one success every 2 seconds (the factor 2 come from the 4 outcome Bell-state measurement of which only 2 amplify the photon, the other two can't be used). 

Let's now explain how photon-amplifiers work in some more details with an explanation based on quantum teleportation with partial entanglement. For this purpose, let's compute what happens when a qubit $\psi=c\ket{0}+s\ket{1}$ (where we assume $c$ real and $s$ complex with $c^2+|s|^2=1$) is teleported using a partially entangled state $\Phi_{pe}=t\ket{0,1}+ir\ket{1,0}$, with t and r real and $t^2+r^2=1$. The form of $\Phi_{pe}$ is chosen to be easily realizable with single-photons and a mere beam-splitter as described below, but for the time being, let's keep it abstract. Let's expand $\psi\otimes\Phi_{pe}$ in the Bell basis for the 2 first qubits with $\phi^\pm=(\ket{00}\pm i\ket{11})/\sqrt{2}$ and $\psi^\pm=(i\ket{01}\pm\ket{10})/\sqrt{2}$ (note the unusual phases $i$ that we chose for later convenience):
\beqa\label{BSM}
\psi\otimes\Phi_{pe}&=&\half\phi^+\otimes(sr\ket{0}+ct\ket{1}) \nonumber\\
&+&\half\phi^-\otimes(-sr\ket{0}+ct\ket{1}) \nonumber\\
&+&\half\psi^+\otimes(cr\ket{0}+st\ket{1}) \nonumber\\
&+&\half\psi^-\otimes(cr\ket{0}-st\ket{1}) \\
&=&ct\ket{001} \nonumber\\
&+&isr\ket{110} \nonumber\\
&+&\half\psi^+\otimes(cr\ket{0}+st\ket{1}) \nonumber\\
&+&\half\psi^-\otimes(cr\ket{0}-st\ket{1}) \label{BSMout}
\eeqa
Let us first concentrate on the case of a Bell-state measurement result $\psi^+$ (third line in eq. (\ref{BSM})). In this case the teleported qubit state reads $cr\ket{0}+st\ket{1}$ which, after normalization -- which corresponds to the probability of that result -- equals $\sqrt{1-|s|^2g^2}\ket{0}+sg\ket{1}$ with a ``gain'' factor 
\beq
g=\frac{t}{\sqrt{c^2r^2+|s|^2t^2}}
\eeq
Fig. 3 illustrates the deformation of the Poincar\'e sphere due to this gain. Note that the gain $g$ depends on the initial state through $c$ and $s$, and that $g\le|s|^2$, so that the probability amplitude of $\ket{0}$ is still real. It is worth rewriting this in terms of the mean value of the z-component, $\eta=\bra{\psi}\sigma_z\ket{\psi}$: 
\beq
\eta \rightarrow \frac{\eta+(2t^2-1)}{1+\eta(2t^2-1)}
\eeq
This expression is identical to the addition of two polarization dependent losses \cite{statPDL}, and also the addition of velocities in special relativity, as in all cases the quantities to be ``added'' are bounded from below and above.

A priori this doesn't look useful, but think of the basic qubit states $\ket{0}$ and $\ket{1}$ as Fock photon number states, i.e. vacuum and single-photon states, respectively. Then the deformation corresponds to an increase of the probability amplitude of the single-photon state by the gain factor $g$. Note that when $t=\sqrt{\half}$, $g=1$, corresponding to standard quantum teleportation with a maximally entangled state. But for $t>\sqrt{\half}$ the gain is larger than 1, i.e. when the heralding signal is on, the probability of an outcoming photon is larger than the probability of an incoming photon: we "amplified the photon".

\begin{figure}
\centering
\includegraphics[width=9cm]{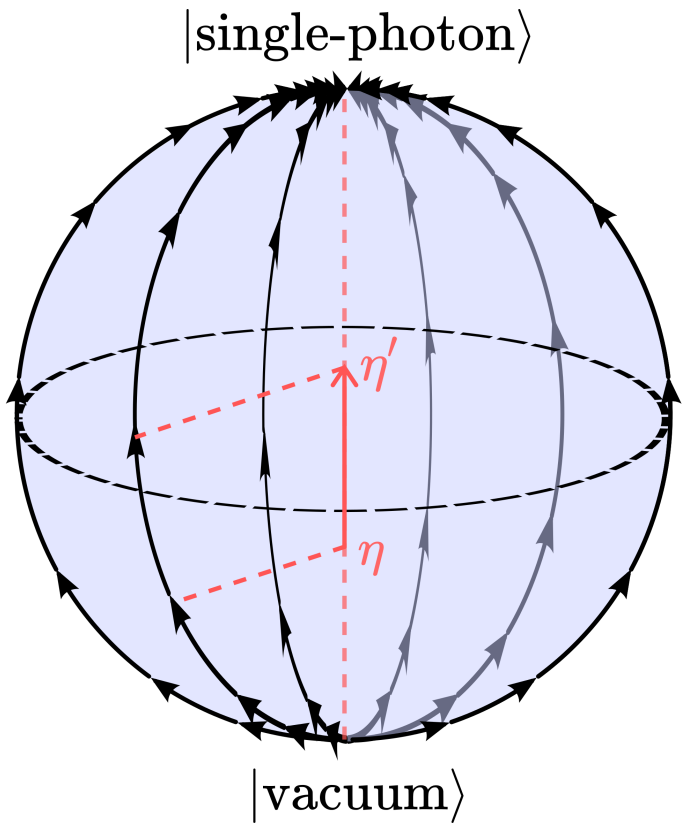}
\caption{Deformation of the Poincar\'e sphere in a photon amplifier. Each point of the sphere is mapped to another point on the sphere (e.g. pure states to pure states), but higher on the z-axis, i.e. closer to the single-photon state. The points on the z-axis move straight up; other points inside the sphere follow more complex trajectories.}\label{fig3}
\end{figure}

Whenever the Bell-state measurement result is $\psi^-$, the analysis is identical, with the same gain factor, except that state $\ket{1}$ gets a $\pi$ phase shift, see last line of eq. (\ref{BSM}). The two other Bell-state measurements results, $\ket{\phi^\pm}$, corresponding to the two first lines of eq. (\ref{BSM}) are also interesting, but we shall not analyze them here as there the phases of the vacuum and single-photon get swapped. Hence, only 2 out of the 4 Bell-state measurement results are relevant. 

Note that in practice one uses threshold single-photon detectors that don't distinguish cases with one or more photons. Equation (\ref{BSMout}) shows what happens: the first line correspond to a result which should never happen (except due to dark counts), as there are no photons arriving on the heralding detectors; the second line should be eliminated using Photon Number Resolving (PNR) detectors, but with non-PRN detectors this results is a too large probability of heralding vacuum. For this reason, when using non-PNR detectors the exact threshold value for $t$ (i.e. the value above which amplification happens) depends on the input state, see \cite{OsorioQubitAmp,AnthonyQubitAmp}.

So far we only considered "photon amplifiers". But one is truly interested in "qubit amplifiers", i.e. devices that amplify the probability amplitude that a photonic qubit (e.g. a polarization or time-bin qubit) is increased. Since the photon amplifier is a coherent quantum process, it can straightforwardly be generalized to qubit amplification by merely first separating the two (polarization or time-bin) modes, next amplify each mode and finally recombining the two modes \cite{qubitamplNG, Curty:2011aa, Pitkanen2011}.

Let us emphasize that in teleportation one always uses entanglement twice. A first time as "quantum teleportation channel" and a second time as results of the Bell-state measurement. This second usage of entanglement for the joint measurement of two systems, typically two qubits, so far received much less attention than entanglement as joint states of distant systems. In particular we have no abstract model for it, contrary to entangled states whose abstract essence is captured by the so-called non-local box of Popescu and Rohrlich \cite{PRbox}.

In summary, qubit amplifiers allow one to increase the probability amplitude of the presence of photonic qubits. The process is probabilistic, but comes with an electronic signal that heralds successful processes. The success probability is at most the probability of presence of a photon in the input port. Hence, this process can't allow one to increase the rate of quantum channels, quite the opposite as the success rate of the amplifier is usually rather low, but it has the great advantage for DIQIP applications that one knows with high confidence when a photon is there, hence one knows when to ask questions to the photon. This is crucial for proper violations of Bell's inequality, as each instance in which Alice and Bob ask a question to their qubit has to be counted in the statistical analysis.

Today, there are only very few experimental demonstrations of qubit amplifiers \cite{NataliaQubitAmp, Kocsis2013}. But the result in \cite{NataliaQubitAmp} allows optimism for the realization of an all fiber practical photonic qubit amplifier at telecom wavelengths. Still, there is plenty of room for much progress.

Finally, Fig. 4 and the caption present the inside of a qubit amplifier. 

\begin{figure}
\centering
\includegraphics[width=9cm]{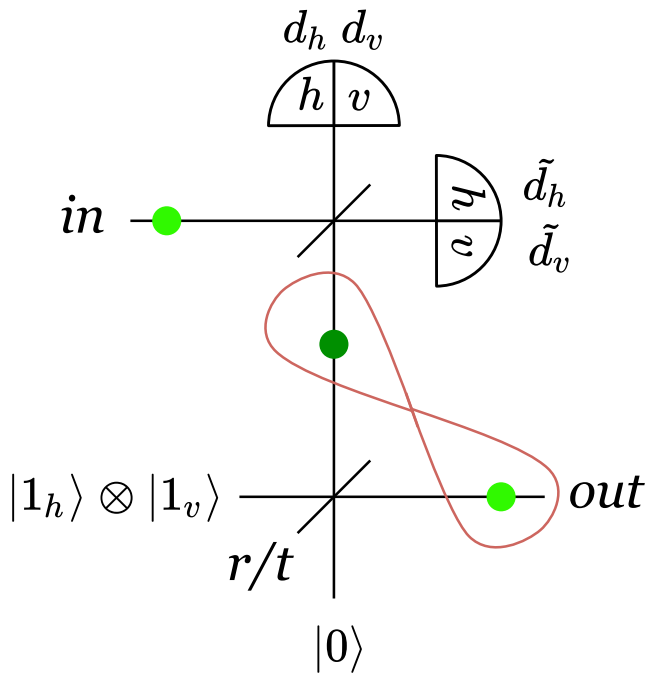}
\caption{A qubit amplifier implementation must contain 2 single-photon sources, one in state $\ket{1_h}$ and the other in state $\ket{1_v}$, for horizontal and vertical polarization. Each photon passes through an imbalanced beam-splitter with transmission t so that each of the 2 photons produce a partially entangled state. The reflected mode meets the in-photonic qubit on a 50-50\% beam splitter, followed by two detectors that herald successful amplification. Each detector distinguishes V and H polarized photons. Ideally, these detectors should be Photon Number Resolving (PNR), so that state $\phi^\pm$ can be discriminated and disregarded. In practice, however, standard non PNR detectors can be used as in the interesting case the probabilities of $\phi^\pm$ is very low (the probability of a in-photon $|s|^2$ is very low).}\label{fig4}
\end{figure}

\section{Conclusion}
In summary, 
\begin{enumerate}
\item Photons can be sent over astronomical distances despite the enormous loss in transmission, as it suffices to send correspondingly enormously large numbers of photons to guarantee that statistically at least some make the entire journey. 
\item	On Earth, optical signals can be sent all around the world thanks to stimulated emission. But no “individual” photon (whatever that means) makes the entire journey.
\item	Direct communication of a bit of information, encoded in a single-photon, is limited to about 500 km, even in the best optical fibers.
\item	Quantum repeaters allow one to overcome this limit, in principle. But the question then is {\it For how long can one store a photonic qubit in a quantum memory?} Roughly the distance is given by the memory time multiplied by the speed of light. This should soon get close to a second for read-write quantum memories.
\item	Heralded probabilistic photon amplifiers (that amplify the probability amplitude of the presence of the photon) are useful to extend quantum communication to Device-Independent applications, i.e. to applications that require that the receiver knows when there is a large chance that a photon is present before choosing in which basis to measure it.
\end{enumerate}

Quantum communication all over our planet is a not too distant dream. Quantum repeaters and high fidelity quantum teleportation will make this possible, though the engineering challenges are still formidable.

\small{
\section*{Acknowledgment} It is a pleasure to thank Hugo Zbinden, Rob Thew and Natalia Bruno for useful comments and help with the figures. Financial support by the Swiss NCCR-QSIT and the European SIQS projects are gratefully acknowledged.
}

\end{document}